\documentclass[aps,amsmath,amssymb, pre,twocolumn]{revtex4}
\usepackage{amsfonts}
\usepackage{bbold}
\usepackage{mathrsfs}
\usepackage{color}
\usepackage{epstopdf}

\usepackage{graphicx}
\usepackage{bm}

\usepackage{natbib}
\usepackage{graphicx}

\usepackage{xcolor}

\usepackage[colorlinks=true,
            linkcolor=blue,
            citecolor=blue,
            urlcolor=blue]{hyperref}

\newcommand{\rr}{{\mathbf r}}

\begin{document}

\title{Macroscopic active matter under confinement: dynamical heterogeneity, bursts, and  glassy behavior in a few-body system of self-propelling camphor surfers}

\author{Marco Leoni}
\thanks{equally contributing authors, listed alphabetically}
\affiliation{Universit\'{e} Paris-Saclay, CNRS, IJCLab, 91405, Orsay, France}

\author{Matteo Paoluzzi}
\thanks{equally contributing authors, listed alphabetically}
\affiliation{Dipartimento di Fisica, Sapienza Universit\`{a} di Roma, Piazzale A. Moro 2, I-00185, Rome, Italy}

\author{Christian Alistair Dumaup}
\affiliation{Department of Physics,  California State University Fullerton, CA 92831 USA}

\author{Farbod Movagharnemati}
\affiliation{Department of Physics,  California State University Fullerton, CA 92831 USA}

\author{Lauren Nguyen-Leon}
\affiliation{Department of Chemistry,  California State University Fullerton, CA 92831 USA}

\author{Tiffany Nguyen}
\affiliation{Department of Physics,  California State University Fullerton, CA 92831 USA}

\author{Sarah Eldeen}
\affiliation{Department of Physics,  California State University Fullerton, CA 92831 USA}

\author{Wylie W.~Ahmed}
\email[correspondence: ]{wylie.ahmed@utoulouse.fr}
\affiliation{Universit\'e de Toulouse, CNRS, Laboratoire de Physique Théorique, Toulouse, France}
\affiliation{Universit\'e de Toulouse, CNRS, Centre de Biologie Int\'{e}grative, Toulouse, France}
\affiliation{Department of Physics,  California State University Fullerton, CA 92831 USA}

\begin{abstract}
We study a few-body system composed of self-propelling camphor surfers confined within a circular boundary. These millimeter-sized particles move in a regime where inertia and long-ranged interactions play a significant role, leading to surprisingly complex and subtle collective dynamics. These dynamics include self-organized bursts and glassy behavior at intermediate densities—phenomena not apparent from ensemble-averaged steady-state measures. By analyzing quantities like the overlap order parameter, we observe that the system exhibits dynamical slowing down as particle density increases. This slowdown is also reflected in the bursting activity, where both the amplitude and frequency of bursts decrease with increasing particle density. A minimal inertial active-particle model reproduces these dynamical steady states, revealing the importance of a new intermediate length scale—larger than the particle size. This intermediate scale is critical for the formation of structures resembling caging and plays a key role in the glass-like transition. Our results describe a macroscopic analog of an active glass with the additional phenomena of density-dependent bursting. 
\end{abstract}

\maketitle

\section{Introduction}
Active matter encompasses systems composed of self-driven entities that exhibit complex and often unexpected behaviors. These systems range from macroscopic biological examples, such as flocks of birds and schools of fish, where inertia plays a significant role~\cite{cavagna2014bird, chate2020dry}, to microscale systems in soft matter, where interactions among particles are comparable to thermal fluctuations~\cite{marchetti2013hydrodynamics,PhysRevX.12.010501,RevModPhys.88.045006, cates2015motility}. At high densities, microscopic active systems, including synthetic and living matter~\cite{angelini2011glass,nishizawa2017universal, parry2014bacterial, klongvessa2019active}, can display glass-like dynamical slowing, yielding, and heterogeneous motion, characteristic of active glasses~\cite{nandi2018random, janssen2019active, mandal2020extreme, goswami2025yielding}. Our focus lies on an intermediate scale --- where  both inertia and non-thermal fluctuations play an important role.

In this study, we explore a few-body system of self-propelling camphor surfers confined within a circular boundary~\cite{soh2008dynamic, nakata2015physicochemical}. These millimeter-sized particles have long-ranged interactions leading to rich and nuanced single-particle~\cite{Camphor2020, boniface2019self} and collective dynamics~\cite{gouiller2021two}.  They have been studied as active particles driven by Marangoni flows that can exhibit  oscillations~\cite{koyano2016oscillatory, xu2021oscillatory, sumino2005self}, synchronization~\cite{sharma2020rotational, kohira2001synchronized, nakata2004synchronized}, intermittent bursting~\cite{suematsu2015synchronized, matsuda2019dynamical}  and turbulence~\cite{bourgoin2020kolmogorovian, gouiller2021mixing}. While the underlying physicochemical processes are complex, here we treat the camphor surfers as a generic system of strongly interacting active particles. Accordingly, we focus on minimal models that capture the effective interactions and emergent dynamics, rather than the full chemical kinetics. This coarse-grained approach is intended to help develop a general framework for understanding interacting active systems, including those where such specific physicochemical mechanisms are absent, such as~\cite{cavagna2014bird, chate2020dry, kumar2014flocking, bar2020self}. Despite the apparent simplicity of the isolated particle trajectories, when multiple particles are present we observe complex phenomena such as self-organized bursts and glass-like behavior, which are not immediately evident from traditional ensemble measures like the mean square displacement ($MSD$).

Previous research has addressed the dynamics of single active camphor particles, revealing different dynamical states~\cite{Camphor2020, suematsu2010mode}. Building on this foundation, we now examine the collective behavior of multiple such particles confined in 2D, highlighting how increasing particle density leads to dynamical slowing down, a hallmark of glassy systems. While dynamical slowing has been observed previously~\cite{soh2008dynamic}, we additionally quantify the intermittent bursting dynamics reminiscent of complex systems~\cite{goh2008burstiness} and plastic deformations~\cite{mandal2020extreme}.  The intermittent bursting dynamics have been observed previously~\cite{suematsu2015synchronized, matsuda2019dynamical}, and in our system we find that the burst amplitude and frequency depend on packing fraction.

\begin{figure}[h!]
    \centering
    \includegraphics[width=0.49\textwidth]{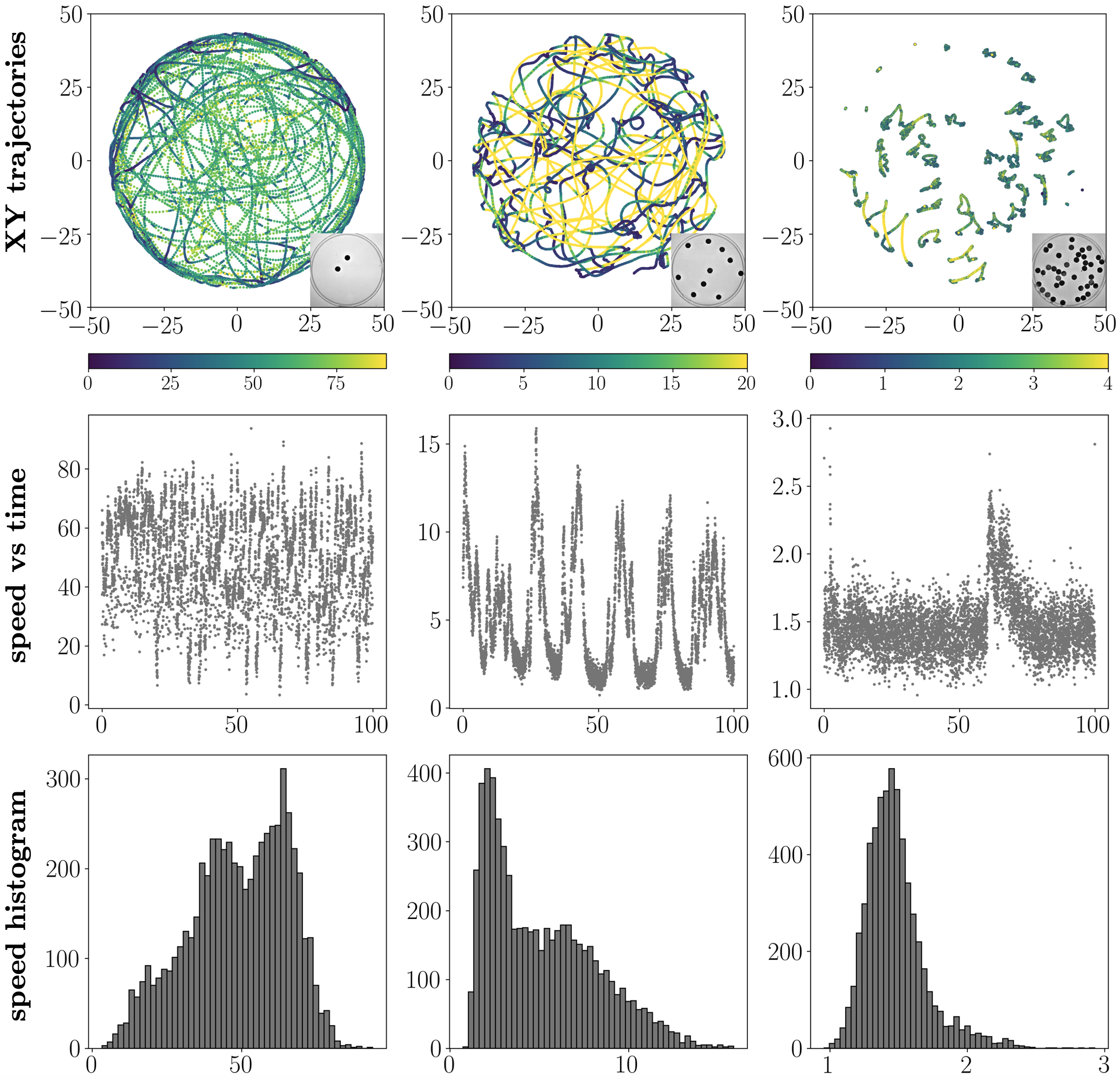}
    \caption{\textbf{Slowing and Bursting Dynamics.} 
    \textbf{Top Row:} XY trajectories (in mm) of individual particles for increasing particle numbers ($\phi = 0.01, 0.06, 0.24$) show the effect of density on motion. The color code represents instantaneous speed (mm/s). At low density ($\phi = 0.01$), particles move freely, whereas at higher densities ($\phi = 0.06, 0.24$), trajectories exhibit increasing confinement and reduced speeds. Insets are representative images. 
    \textbf{Middle Row:} Time series (in seconds) of ensemble-averaged particle speed (mm/s) demonstrate the transition in dynamics with density. At low density ($\phi = 0.01$), the speed is noisy and relatively constant. At intermediate density ($\phi = 0.06$), organized collective bursts dominate, while at high density ($\phi = 0.24$), bursts become less frequent and the overall speed decreases. 
    \textbf{Bottom Row:} Speed histograms (in mm/s), constructed from ensemble-averaged speed above, illustrate the changes in speed distributions across densities. At low density, a broad distribution is observed. As density increases, distributions narrow, indicating a slowdown. Note that different densities use different color scale and axes to highlight local differences.}
    \label{fig:ExpTraj} 
\end{figure}

Dynamical slowing down is a hallmark of glassy systems, where particle motion becomes increasingly sluggish with variation in a control parameter, in this case the particle density. This slowdown can be linked to dynamical heterogeneity, a concept central to the study of glass-forming systems~\cite{janssen2019active, dauchot2005dynamical}. In dense glasses, particles are dynamically trapped in cages formed by their neighbors, leading to a slow relaxation process~\cite{berthier2011theoretical}. Similarly, in our system, we find that as particle density increases, particles are confined within transient “cages” formed by neighboring particles, albeit at much lower densities than typically reported~\cite{berthier2011theoretical}. This confinement eventually leads to collective bursts of motion~\cite{suematsu2015synchronized}, where particles momentarily escape their cages and exhibit a burst of speed, followed by periods of inactivity, reflecting a behavior akin to dynamical heterogeneity in glassy systems but with the added feature of time periodicity (Fig.~\ref{fig:ExpTraj}).

Through experimental observations and a minimal soft-matter model of interacting active particles, we identify the emergence of an intermediate length scale --- larger than the particle size --- in this system, potentially linked to the length scale of chemical gradients driving self-propulsion~\cite{boniface2019self}. This intermediate scale plays a critical role in the formation of cage-like structures at intermediate density, reminiscent of the caging observed in glass transitions. Such length scales have been proposed in active matter systems to explain the complex spatiotemporal organization seen in these systems~\cite{soh2011swarming, araujo2023steering}. This minimal approach is chosen to enable generalization to interacting active systems beyond soft matter.

This study contributes to the broader understanding of glassy dynamics in active matter by illustrating a macroscopic analog, complementing existing studies at the microscale~\cite{hunter2012physics, lu2013colloidal, berthier2016facets}. Specifically, we explore how density influences the collective behavior of self-propelled particles in confinement and how it can drive the system toward a glass-like transition --- a phenomenon that has been difficult to observe in experiments~\cite{janssen2019active}.  Remarkably, the observed phenomena occur at intermediate particle densities, a unique feature of this active macroscopic system.

\section{Experimental system}
Millimeter-scale camphor surfers were created by infusing agarose gel disks with camphor solution as studied previously~\cite{Camphor2020, soh2008dynamic}. 
The resulting self-propelled surfer has a radius of $\sim 3.5$ mm and a mass of $\sim 40$ mg.
The dynamics are studied by placing the surfer at the water-air interface in a circular petri dish of 9 cm diameter with 20 g of ultrapure water.
Self-propulsion is driven by gradients in surface tension~\cite{suematsu2014quantitative}. The active particle is free to move in-plane but experiences a vertical wall at the boundary. The collision with the boundary is likely mediated through capillary effects~\cite{soh2008dynamic}. 

Images were captured using a CMOS camera and lens (Basler acA3088-57$\mu m$ and Computar M3Z1228C-MP, from Graftek Imaging) at 60 Hz where 4$x$ pixel binning was used at the time of acquisition, resulting in an image of 768 $\times$ 516 pixels and saved as individual linearly encoded TIFF files. The total duration of each data acquisition was 100 seconds. Image sequences were analyzed in MATLAB to determine particle trajectories using a custom-written image processing code. Briefly, images were thresholded, and background noise was removed via filtering, and the centroid of the single particle was recorded for each frame. Because the macroscopic particles remain in-plane and exhibit high contrast, their centroids could be reliably identified in every frame. Tracking precision was determined to be $\sim1$ pixel, resulting in an uncertainty of 0.2 mm. This tracking precision corresponds to  $\sim1/30$th of the particle diameter.

  \begin{figure}[h!]
       \centering
\includegraphics[width=0.49\textwidth, trim= 5 0 0 0, clip]{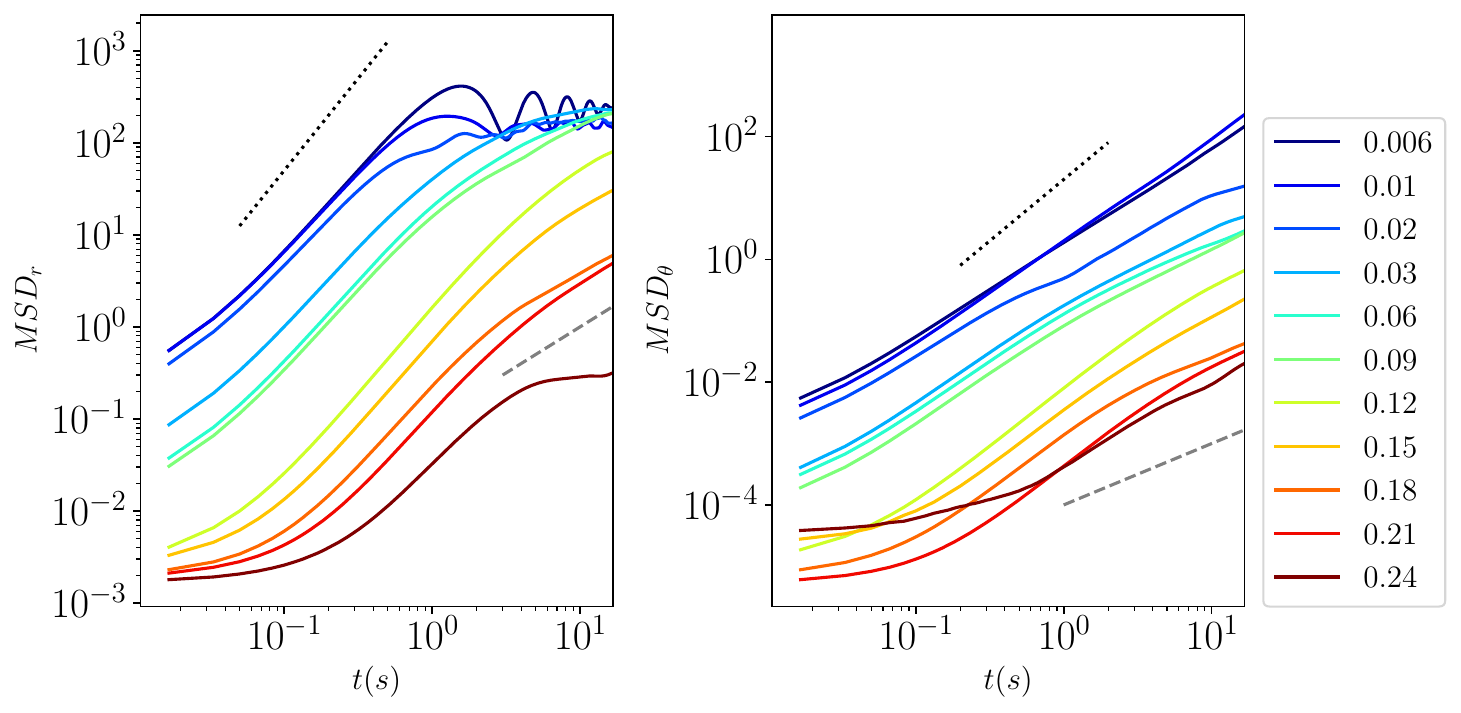} 
       \caption{
       {\bf Mean Squared Displacement:}  
       As packing fraction ($\phi$) increases, the particles systematically exhibit less motion.  At low $\phi$, radial motion is ballistic with a plateau corresponding to container size, and angular motion is persistent.  At intermediate $\phi$, particles become more diffusive in both radial and angular directions.  At high $\phi$, particles exhibit short time scale caging, with diffusive-like behavior at longer times. Dotted and dashed lines are guides for the eye for ballistic ($t^2$) and diffusive ($t$) scaling, respectively.
       }
\label{fig:MSD-new}
   \end{figure}

\section{Main findings}

\subsection{Ballistic, diffusive, and caged motion}

In previous work on isolated particles in confinement, the $MSD$ revealed distinct plateaus and crossovers, highlighting the complex dynamics of single particles~\cite{Camphor2020}. Extending this analysis to a many-particle system, the ensemble averaged angular and radial $MSD$s exhibit density-dependent behavior. As the packing fraction increases, particles experience local caging, slowed dynamics, and a transition from ballistic to diffusive motion.

We computed both the radial and the angular $MSD$ for $X_i = \rr_i, \theta_i$, where $\rr_i$ and $\theta_i$ are conventional polar coordinates, and the $MSD$ is defined as:
\begin{align}
MSD_{X_i} = \frac{1}{N} \left\langle \sum_i \left[ X_i(t) - X_i(0)\right]^2 \right\rangle \; .
\end{align}
where $X_i$ is the coordinate, $t$ is time, and the angle brackets indicate the average over individual trajectories, $i$. The angular, $MSD_\theta$, has been computed using the angular velocity, $\dot{\theta}$. As shown in Fig.~\ref{fig:MSD-new}, at low packing fractions ($\phi < 6 \%$), particles perform long ballistic runs and eventually collide with the boundaries of the container. This corresponds to a ballistic regime in the radial, $MSD_r \sim t^2$, extending up to the container radius. This is consistent with what we observed at the single particle level. As density increases, we observe an attenuated amplitude as shown by a consistent shift downward of the $MSD_r$, apparent caging shown by the low timescale plateau, and a transition in both the radial and angular $MSD$ that changes from ballistic to diffusive, i.e. $MSD \sim  t$, on longer time scales. Moreover, for packing fractions larger than $12 \%$, particles spend most of the time exploring their local area. At the same packing fractions, the $MSD_\theta$ is not ballistic anymore and tends to develop a diffusive regime.

\begin{figure}[h!]
       \centering
\includegraphics[width=0.49\textwidth, trim= 5 0 0 0, clip]{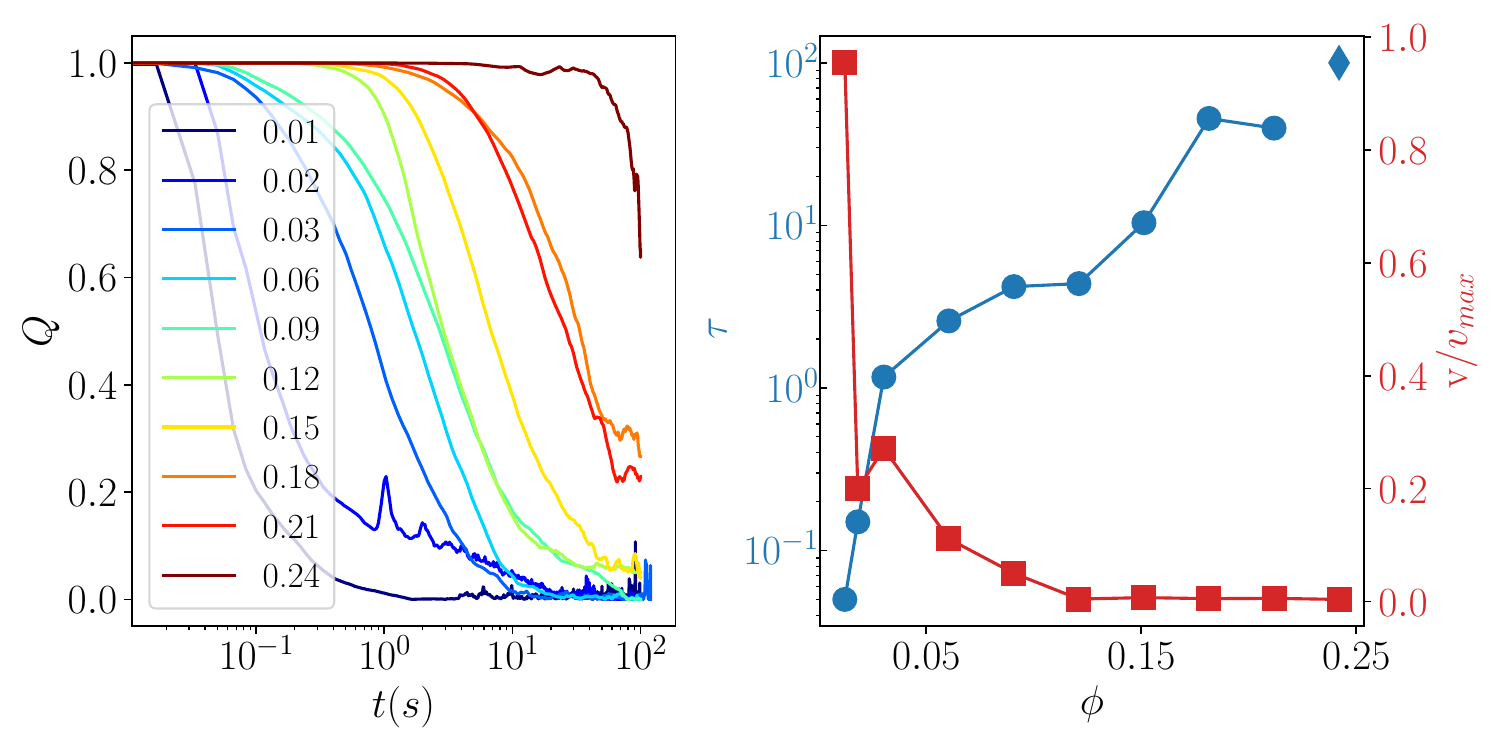}
       \caption{
       {\bf  Overlap parameter, average speed and relaxation time.}
      Left:  Overlap parameter, $Q(t)$, computed from experimental trajectories. Right: The structural relaxation time $\tau_\alpha$ computed through the overlap parameter increases as $\phi$ increases (blue circles). 
     The last data point (diamond)  provides a lower bound obtained from the experimental data, see the main text. 
      This behavior mirrors the decay of the average velocity (red squares) that tends to zero as $\phi$ increases. 
        }
       
    \label{fig:Q(t)}
\end{figure}

\subsection{Average Speed and Slowing Down} 

As indicated by the decreasing amplitude of the $MSD$ (Fig.~\ref{fig:MSD-new}), particles exhibit a pronounced slowing down in both their radial and angular dynamics as the density increases. This slowing behavior points to an evolving structural organization within the system that is not captured in the $MSD$. 

To quantify the extent of structural relaxation, we measure the dynamical overlap parameter \cite{lavcevic2003spatially}, defined as:

\begin{align}
    Q(t) = \frac{1}{N} \left\langle \sum_i \Theta( \delta - |\mathbf{r}_i(t) - \mathbf{r}_i(0)| ) \right\rangle,
    \label{eq:Q}
\end{align}
where \( Q(t) \) represents the fraction of particle displacements less than \(\delta\) at time \(t\) \cite{lavcevic2003spatially}. 
This observable, that is the dynamical counterpart of the Edwards-Anderson order parameter in lattice models, provides a quantitative measure of the structural relaxation time.
Here, we set \(\delta = 2a\), with \(a\) being the particle radius and $\Theta(x)$ is the Heaviside step function. A non-zero value \(q = \lim_{t \to \infty} Q(t) \neq 0\) signals an ergodicity breaking in the system, indicating that a fraction \(q\) of particles remain trapped in their local environment (Fig.~\ref{fig:Q(t)}).

From \(Q(t)\), we define the structural relaxation time \(\tau_\alpha\) as the time at which \( Q(\tau_\alpha) = e^{-1} \). The structural relaxation time \(\tau_\alpha\) serves as an estimator for the emergence of complex dynamics, where local particle rearrangements are only possible through cooperative mechanisms, such as particles overcoming caging by their neighbors. A hallmark of glassy dynamics is the rapid growth of \(\tau_\alpha\) with increasing control parameter, here the particle density. This behavior, shown in Fig.~\ref{fig:Q(t)}, is a clear fingerprint of glassy dynamics in our active system.
The last data point (diamond) of  Fig.~\ref{fig:Q(t)} instead 
provides a lower bound $\tau^*_\alpha$  
obtained as
$ Q(\tau^*_\alpha) = Q_{m} $ where  $ Q_{m}$ is the minimum value of $Q$ observed experimentally for  $\phi=0.24$, dark red curve  on the left panel of Fig.~\ref{fig:Q(t)}.

\begin{figure}[h!]
       \centering
\includegraphics[width=0.49\textwidth, trim=80 20 70 40, clip]{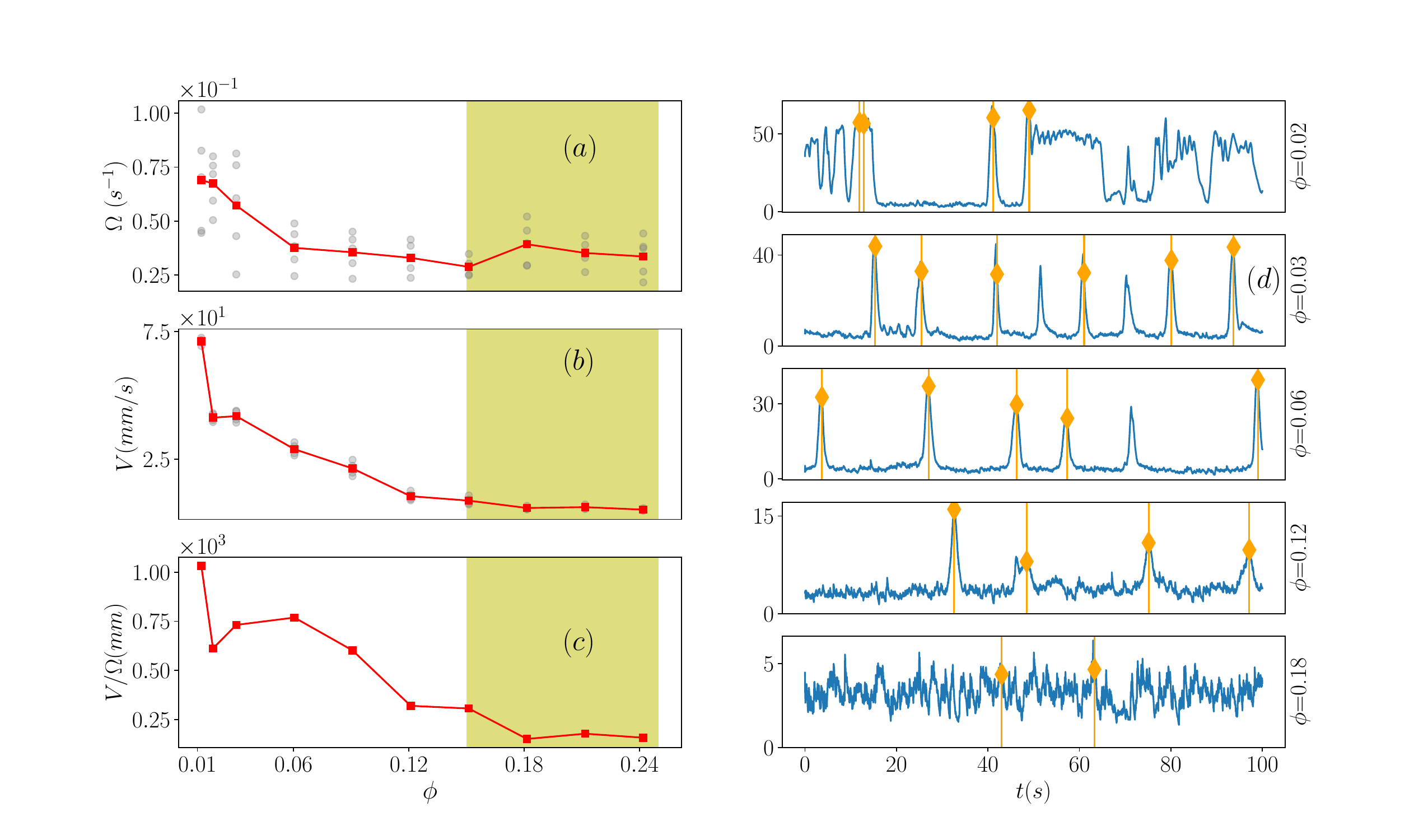}
      \caption{
\textbf{Bursting Behavior and Slowing Down
} 
(a) Frequency and (b) amplitude of bursts as functions of the particle density, showing an overall decreasing trend with increasing density. Bursts are defined as peaks in the squared-speed signal exceeding a given threshold \(p\) (see the main text); different realizations (gray symbols) are averaged (red line). The shaded region in (a-c) indicates where bursts become rare, leading to limited statistics. 
(c) Burst breadth: typical length scale $L$ associated to a burst, obtained by combining the burst amplitude 
$V$ (dimensionally a velocity) of panel (b) 
and the burst frequency $\Omega$  of panel (a)
 as $L = V/\Omega$. This plot indicates that the typical displacements of a collection of particles during a burst event is locally maximized at intermediate density($\phi=0.06$). The burst breadth decreases at increasing density. 
(d) Speed (in mm/s) of an individual particle at varying packing fractions: from low (top) to high (bottom). 
At low packing fractions, higher peaks (corresponding to higher speeds) are observed, while peaks 
diminish and intervals between them increase at higher densities, reflecting particle slowdown. Yellow diamonds and vertical lines indicate the burst amplitude and time, respectively, in randomly chosen trajectories.
}

\label{fig:Bursts-Many}
   \end{figure}

\subsection{Bursting Behavior in Particle Squared Speed} 

Particles exhibit intermittent, abrupt changes in their motion, which we term ``bursting'' behavior. By examining the speed of individual particles, we observe distinct bursts. 
While similar bursting behavior has been reported in both single~\cite{suematsu2010mode, xu2021oscillatory} and multiparticle systems~\cite{suematsu2015synchronized, matsuda2019dynamical}, here we examine how the collective dynamics depend on particle density.

We quantify bursting by identifying peaks in the squared-speed time series that exceed a threshold set by the quantile \(p\) of the distribution. Because the signal is strongly non-Gaussian, a quantile-based criterion is more robust than one based on the mean. In practice, we use high quantiles (\(p=0.9\)–\(0.98\)), since lower thresholds are overly sensitive to variability in peak height. For each particle and each experiment, the squared-speed trace is first smoothed (using a rolling average over $\sim$0.5 s) to reduce noise. Values below the chosen threshold are then discarded, and peaks are identified from the remaining signal by isolating the positive-slope segments associated with burst onset. The burst amplitude is defined as the square root of the squared-speed value at each detected peak, while the burst frequency is obtained from the mean interval between successive peaks, \(\bar{T}=\frac{1}{N}\sum_{n=1}^{N}T_n\), with \(\Omega = \bar{T}^{-1}\).


The amplitude of these bursts decreases as the particle density increases~(Fig.~\ref{fig:Bursts-Many}b), while the time intervals between consecutive bursts become longer~(Fig.~\ref{fig:Bursts-Many}a). This trend reflects the overall slowing down of particle motion with increasing density. Further, the burst breadth, defined as $V / \Omega$ provides a characteristic length scale of the bursts which shows a local peak at $\phi = 0.06$~(Fig.~\ref{fig:Bursts-Many}c). Looking at individual trajectories of representative particles, as shown in Fig.~\ref{fig:Bursts-Many}d, it is clear that intermittent bursting emerges at intermediate densities and the burst frequency decreases with increasing $\phi$.

This intermittent bursting regime is observed at intermediate densities, reflecting a distinctive interplay between caging, activity, and collective motion. While bursting has been observed in related systems~\cite{suematsu2015synchronized, dey2025confinement}, here we quantify it alongside glass-like slowing with increasing density, linking these dynamics to active glass behavior with long-ranged interactions.

At higher densities, the aperiodic bursting resembles dynamical heterogeneity in glasses, where particle dynamics are non-uniform in space and characterized by localized groups of collectively rearranging particles, while the rest of the system remains temporarily frozen. This behavior, often attributed to caging effects, necessitates increasingly large cooperative motion to mobilize particles as density increases~\cite{janssen2019active}. It is interesting to note that, a similar slowing of bursting with increasing density was observed in a dry active system~\cite{patterson2017clogging}, suggesting this phenomenon may be observable across a wide range of physical systems.

To understand bursting dynamics in marangoni-driven particle systems, recent studies have developed detailed models based on the underlying physicochemical mechanisms, including reaction–diffusion frameworks~\cite{matsuda2019dynamical}, bistable chemical potentials~\cite{dey2025confinement}, and coupled chemical fields with hydrodynamic inertia~\cite{foradori2026inertial}. In the following sections, we instead adopt a simpler, minimal modeling approach aimed at capturing the effective interactions and emergent dynamics, allowing generalization beyond specific physicochemical realizations to broader classes of interacting active systems.

\section{An analytic model for bursting}

The experimentally observed decrease in burst frequency with increasing particle density (Fig.~\ref{fig:Bursts-Many}a) is a nontrivial feature of the system.  Most studies of bursting coupled oscillators study phase synchronization~\cite{rosenblum1996phase, ivanchenko2004phase},  while frequency remains relatively constant due to an inherent natural frequency set by individual oscillators. In our case, an individual particle has no inherent oscillation frequency, the bursts collectively emerge due to multi-particle interactions, and their frequency decreases with $\phi$.
To explore whether this trend can emerge from basic principles, we analyze a minimal model of hydrodynamically coupled active oscillators. Rather than focusing on phase synchronization (as in Kuramoto-like models~\cite{rosenblum2001phase}), our goal is to understand how the timescale of bursts—here modeled as oscillations—depends on density.

\begin{figure}
\includegraphics[width=0.4\textwidth]{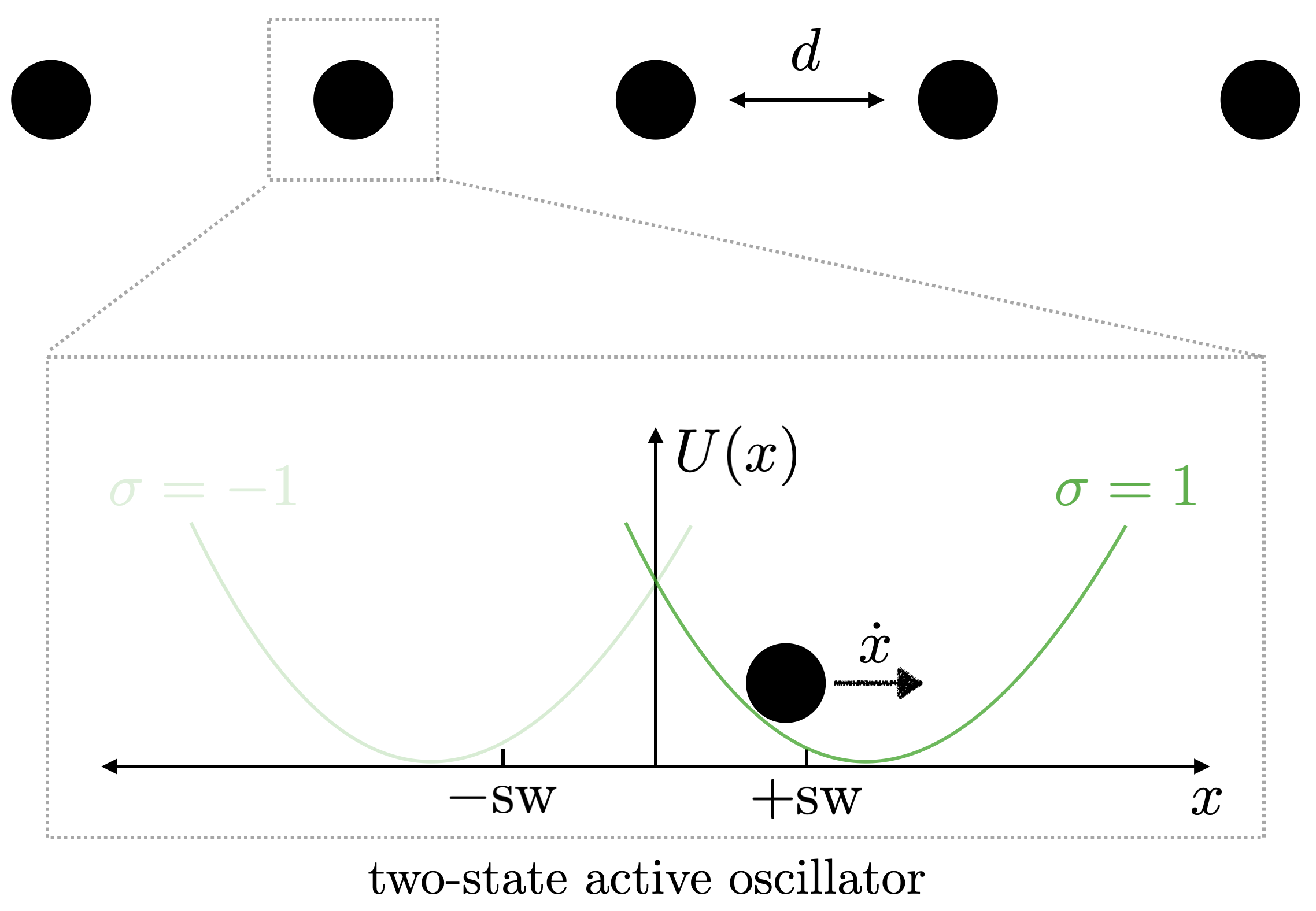}
\caption{\textbf{Analytic Model of Slowing Bursts.} A minimal model of hydrodynamically coupled two-state oscillators. Each particle moves between positions ±sw under a harmonic potential $U(x)$ that switches sign at the turning points. Coupling is introduced via long-range hydrodynamic interactions that scale as $1/d$, where $d$ is the interparticle distance. 
}
    \label{fig:Rower}
\end{figure}

We consider a series of identical pointlike particles (Fig.~\ref{fig:Rower}), each described as a two-state active oscillator, interacting only through long-range hydrodynamic coupling~\cite{leoni2009basic, kotar2010hydrodynamic, Lagomarsino2003}. Each oscillator moves in a piecewise harmonic potential $U(x)$ that reverses sign at the turning points. Coupling is introduced via long-range hydrodynamic interactions that scale as $1/d$, where $d$ is the interparticle distance.
Hence, the effective density is $\rho \sim 1/d$.

When several oscillators are hydrodynamically coupled, the equations of motion for any given particle $n$ with $n=1, \ldots N$, are:
\begin{equation}
\gamma \dot{x}_n = -k(x_n - \sigma_n\, \mathrm{sw}) - k \gamma \sum_{m \neq n } H_{nm}(x_m - \sigma_m\, \mathrm{sw})
\end{equation}

For spherical particles of radius $a$, 
$\gamma = 6 \pi \xi  a$  is the Stokes' coefficient where  $\xi$ is the  viscosity 
of the surrounding fluid \cite{Lagomarsino2003}
and $H_{nm} = 1/(4 \pi \xi |x_n - x_m|)$ represents the hydrodynamic interaction between two particles $m, n$ with $m\neq n$. 
Focusing on the case $N=2$ and considering the anti-phase stationary solution (where $x_1 = -x_2$ and $\sigma_1 = -\sigma_2$), we find that the effective frequency of oscillation becomes:
\begin{equation}
\omega = k(1/\gamma - H_{12}) \sim \omega_0(1 - \beta \rho).
\end{equation}
where $\beta = 3 a/4\pi $
and $\omega_0 = k/\gamma $ corresponds to the oscillation frequency of a single, non-interacting, rower particle.
Thus, the oscillation frequency decreases with increasing density. This result generalizes to $N$ particles lying on a line at fixed distance $d$ and moving in alternating anti-phase, yielding:
\begin{equation}
\omega = k\left[1 + \beta \rho \sum_{m=1}^{N-1} \frac{(-1)^m}{m} \right].
\end{equation}
In the limit $N\to \infty$, the
alternating harmonic series converges to a negative constant, ensuring that $\omega$ decreases with $\rho$ even in larger assemblies.

This simple model shows that long-ranged interactions among active oscillators can in some circumstances, e.~g. the antiphase motion of the above example, lead to decreasing oscillation frequency as density increases. The slowing of coordinated motion in this minimal setting mirrors the experimental observation of burst frequency reduction with increasing particle number. While the real system involves richer dynamics in 2D and nonlinearities, this toy model illustrates that hydrodynamic coupling alone can account for a key aspect of bursting behavior, i.~e., slowing down with increased particle density.

\section{Numerical simulations}

\begin{figure}[h!]
       \centering
\includegraphics[width=0.49\textwidth, trim= 5 0 0 0, clip]{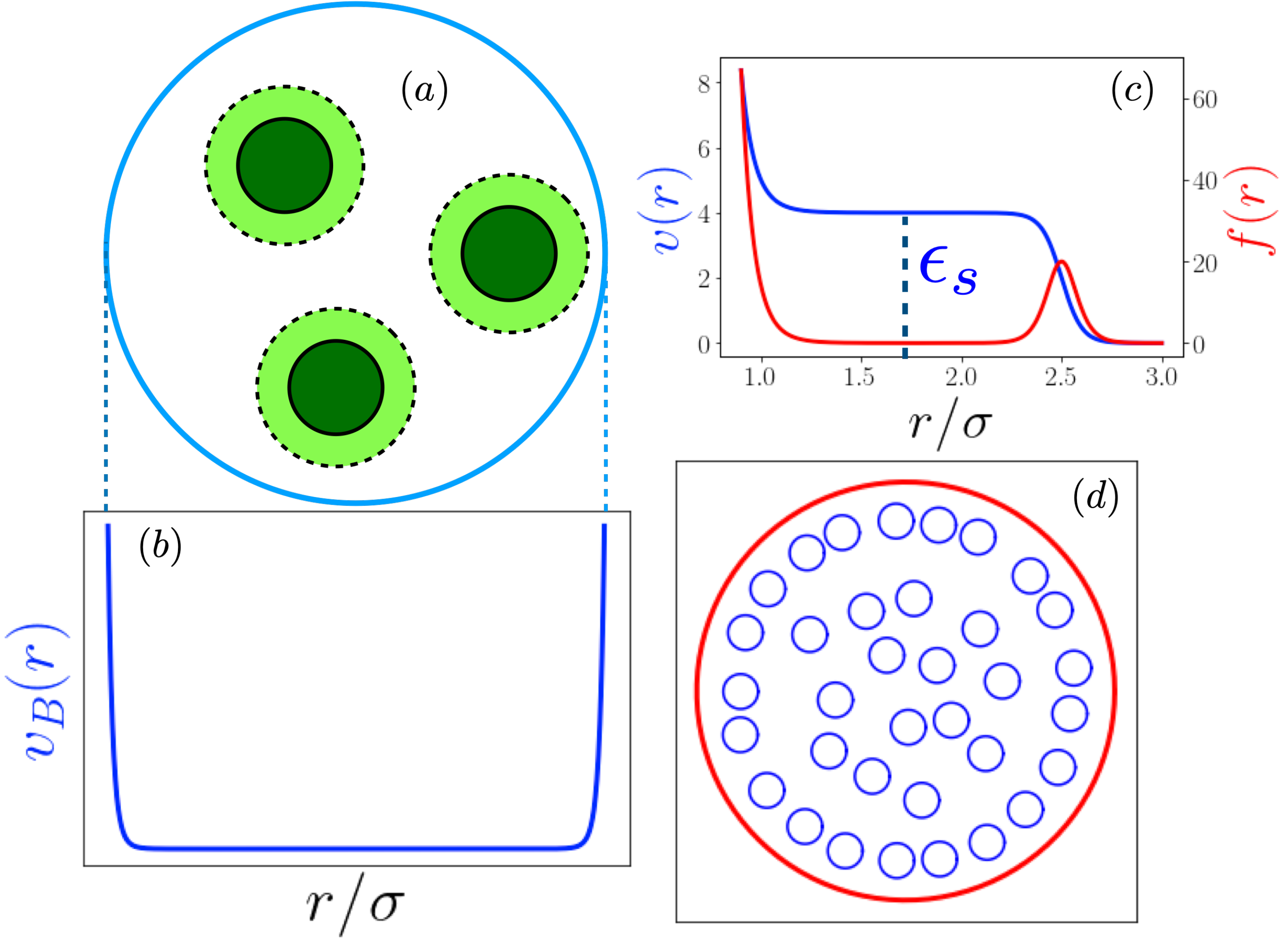}
       \caption{ {\bf Numerical Model.}
(a) Schematic of three camphor-like particles confined in a circular container. Dark green shows the short-range excluded-volume interaction, modeled with a steep $r^{-12}$ repulsion. Light green indicates a second, softer repulsive length scale used to mimic long-ranged interactions.  
(b) Effective repulsive potential used to model the confining boundary.  
(c) Pair potential with two repulsive length scales (blue) and the corresponding force (red).  
(d) Example of a stationary configuration with $N=35$ active particles confined in a circular box of radius $R=6\sigma$, where $\sigma$ denotes the particle diameter.}
\label{fig:Model}
   \end{figure}

In experiments, we observe a clear increase in the structural relaxation time of the system, as quantified by the decay of 
$Q(t)$, when the number of camphor surfers is increased. From a theoretical perspective, this behavior signals a dynamical slowing down characteristic of glassy systems. However, in contrast to colloidal glasses, the packing fractions at which this phenomenology emerges are extremely small. To gain insight into this puzzling observation, we developed a simple numerical model that reproduces this feature. We simulate a minimal model of confined active particles that retains the ingredients we deem essential: inertia \cite{lowen2020inertial,caprini2021inertial}, hard-wall confinement, excluded-volume interactions, and a second, longer-range repulsive length scale. We ask whether such a system can display glass-like slowing at moderate packing fractions; it does, as demonstrated below. Particles interact via a two–length-scale pair potential (Fig.~\ref{fig:Model}), comprising a steep short-range steric repulsion and a soft shoulder at a larger distance (schematized in light green in Fig.~\ref{fig:Model}) that encodes gradual, long-range repulsion.

Millimeter-scale active particles are intrinsically complex, extended objects. For simplicity, we approximate each particle as a solid disk, retaining only the minimal features required to capture the observed dynamics. The system is therefore modeled as $N$ Active Brownian Particles in two spatial dimensions, evolving in the underdamped regime.
\begin{align}
    m \ddot{\mathbf{r}}_i &= -\gamma \dot{\mathbf{r}}_i + u_0 \mathbf{e}_i  + \mathbf{f}_i + \mathbf{f}^B_{i}\\ 
    I \ddot{\theta}_i &= -\Gamma_\theta \dot{\theta}_i + \eta_i
\end{align}
Here, $\mathbf{r}_i$ is the position of particle $i$ ($i=1,\dots,N$), and $\theta_i$ is the angle that sets the direction of self-propulsion, with $\mathbf{e}_i = (\cos \theta_i, \sin \theta_i)$. We choose units such that $\Gamma_\theta = \gamma = m = I = 1$, ensuring that inertia remains important. The term $\eta_i$ represents Gaussian noise characterized by $\langle \eta_i(t) \rangle =0$, and $\langle \eta_i(t) \eta_j(s) \rangle = 2 \tau^{-1} \delta_{ij} \delta(t-s)$ that drives rotational diffusion of the propulsion direction on a timescale $\tau$, where $\tau = u_0 = 1$. In this inertial regime, the persistence length of the particle trajectories is larger than the standard value $u_0 \tau$.

The term $\mathbf{f}^B_i$ represents the force from the confining boundary. This force has been modeled using a repulsive potential $v_B(r) = (\sigma / r)^{12}$ (as depicted in Fig.~\ref{fig:Model}(a,b)), with $\sigma=1$.
The second force term, $\mathbf{f}_i$, accounts for pairwise interactions between particles inside the container.

\begin{align}
    \mathbf{f}_i = \sum_{j\neq i} -\frac{dv(r_{ij})}{d r_{ij}} \frac{\mathbf{r}_{ij}}{r_{ij}}
\end{align}
Here, $\mathbf{r}_{ij} = \mathbf{r}_i - \mathbf{r}_j$ is the relative position and $r_{ij} = |\mathbf{r}_{ij}|$ its magnitude. The pair interaction $v(r)$ includes two repulsive length scales, described by a repulsive-shoulder potential~\cite{PhysRevE.79.051202,martin2022dynamical}, as illustrated in Fig.~\ref{fig:Model}(c).
\begin{align}
    v(r) = \left( \frac{\sigma}{r} \right)^n + \frac{\epsilon_s}{2} \left[ 1 - \tanh{ k \left( r - \sigma_s \right)} \right] \; .
\end{align}

Previous work has shown that introducing a second repulsive length scale can drive a glass transition in overdamped Active Brownian particles~\cite{martin2022dynamical}. Our model adds further complexity by including (i) inertia in both translational and rotational motion and (ii) strong confinement. In the simulations, we set the particle size as the unit length ($a=1$). The potential parameters are $n=14$, $\sigma_s=1.5$, and $k=1$. Because the container radius is only a few particle diameters, crystallization is likely. To prevent this, a common strategy in numerical simulations is to introduce geometrical frustration that avoids crystallization.
We thus introduce polydispersity by sampling $\sigma_s$ from a power-law distribution, $P(\sigma_s) \propto \sigma_s^{-3}$, within the interval $[0.7\sigma_s, 1.4\sigma_s]$.
Polydispersity of the particle size is surely present in the experiment as well, e.g., due to imperfections in particle fabrication. The fact that the phenomenology does not depend on the specific shape of the distribution \cite{PhysRevX.7.021039}, allows us to dispense with a precise experimental determination of the particle-size distribution.

\begin{figure}[h!]
       \centering
\includegraphics[width=0.49\textwidth, trim= 5 0 0 0, clip]{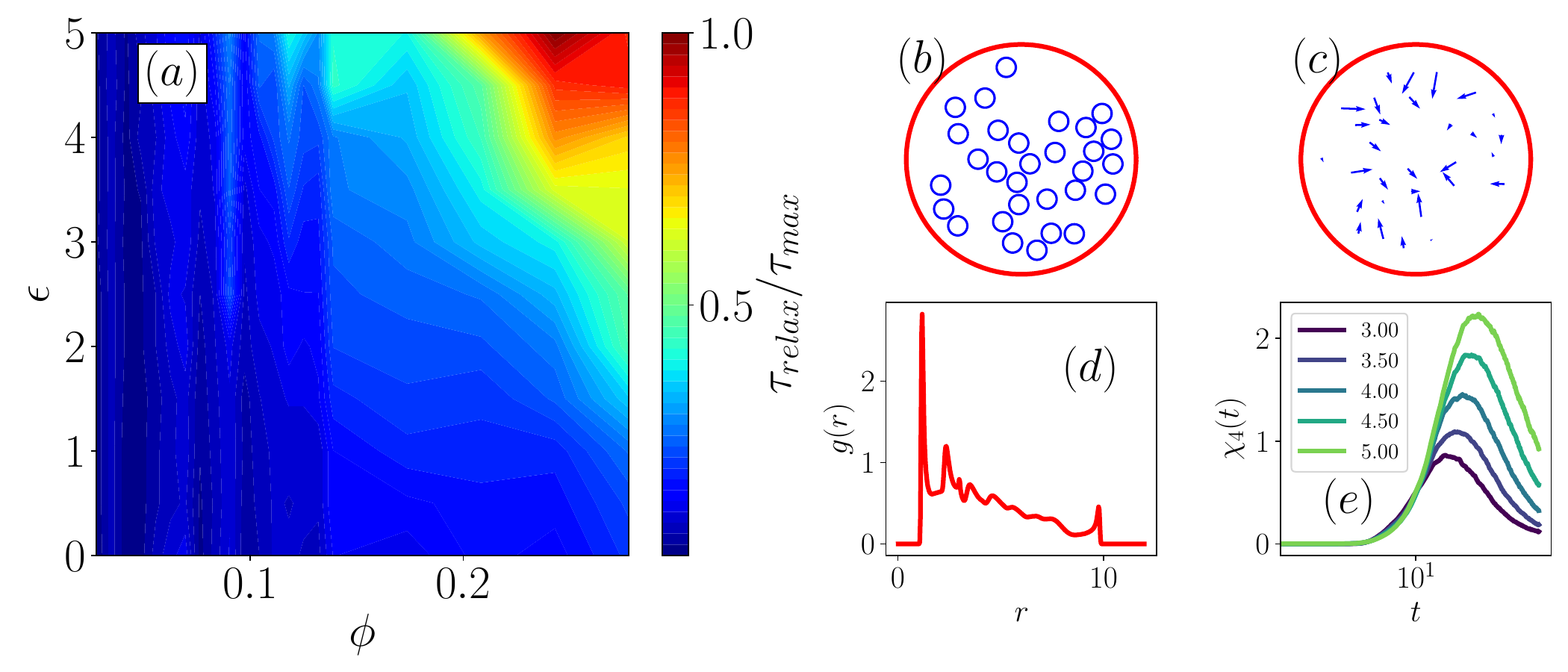}
       \caption{ {\bf Phase Diagram.}
        (a) Phase diagram of the model using the packing fraction $\phi$ and the strength of the second repulsive potential $\epsilon$ as control parameters. The color map indicates the typical relaxation time $\tau_\alpha$. The region in red (large $\epsilon$) represents where the system undergoes a dynamical slowing down.
        (b) Typical stationary snapshot for $\epsilon=5.0$. (c) Map of displacement computed over $\tau_\alpha$ for $\phi =0.24$ and $\epsilon=0.5$.
(d) Radial distribution function $g(r)$ for $\epsilon=5$ and $\phi=0.28$.
(e) Dynamical susceptibility $\chi_4(t)$ as $\epsilon$ increases (from violet to light green, see legend).} 
\label{fig:Numerics}
   \end{figure}

We first mapped the phase diagram of the model by varying the strength of the second repulsive scale, $\epsilon_s$, and the packing fraction, $\phi$. To characterize the system, we measured positional order using the radial distribution function $g(r)$ and dynamics using the structural relaxation time $\tau_\alpha$ extracted from the overlap parameter $Q(t)$. The absence of sharp peaks in $g(r)$ confirms that crystallization is avoided due to particle polydispersity  (Fig.~\ref{fig:Numerics}(d) shows a typical $g(r)$). The phase diagram in Fig.~\ref{fig:Numerics}(a) shows $\tau_\alpha$ as a color map, revealing a region of pronounced dynamical slowing down for large repulsion strengths ($\epsilon_s > 2$). A typical stationary configuration at $\phi=0.24$ is shown in Fig.~\ref{fig:Numerics}(b). Displacement maps highlight heterogeneous particle motion, especially at $\epsilon_s = 5.0$ (Fig.~\ref{fig:Numerics}(c)), where arrows represent displacements $\Delta \mathbf{r}_i(\tau_\alpha) = \mathbf{r}_i(\tau_\alpha) - \mathbf{r}_i(t_0)$, with $t_0$ taken once the system is stationary. For fixed $\epsilon_s$, this glassy behavior becomes stronger as $\phi$ increases, in line with our experimental observations. This is confirmed by the behavior of the dynamical susceptibility $\chi_4(t)$, which provides a quantitative measure of dynamical heterogeneity (see ~\cite{lavcevic2003spatially} for details). 
$\chi_4(t)$ is a dynamical susceptibility that quantifies the presence of competing 
relaxation times in the collective dynamics. A broad shape of $\chi_4(t)$ indicates that cooperative relaxation processes take place, as those peculiar of glassy systems. 
The broader shape of $\chi_4(t)$ signals the presence of dynamical heterogeneity as the strength of the second repulsive length scale increases.

\begin{figure}[h!]
       \centering
\includegraphics[width=0.5\textwidth]{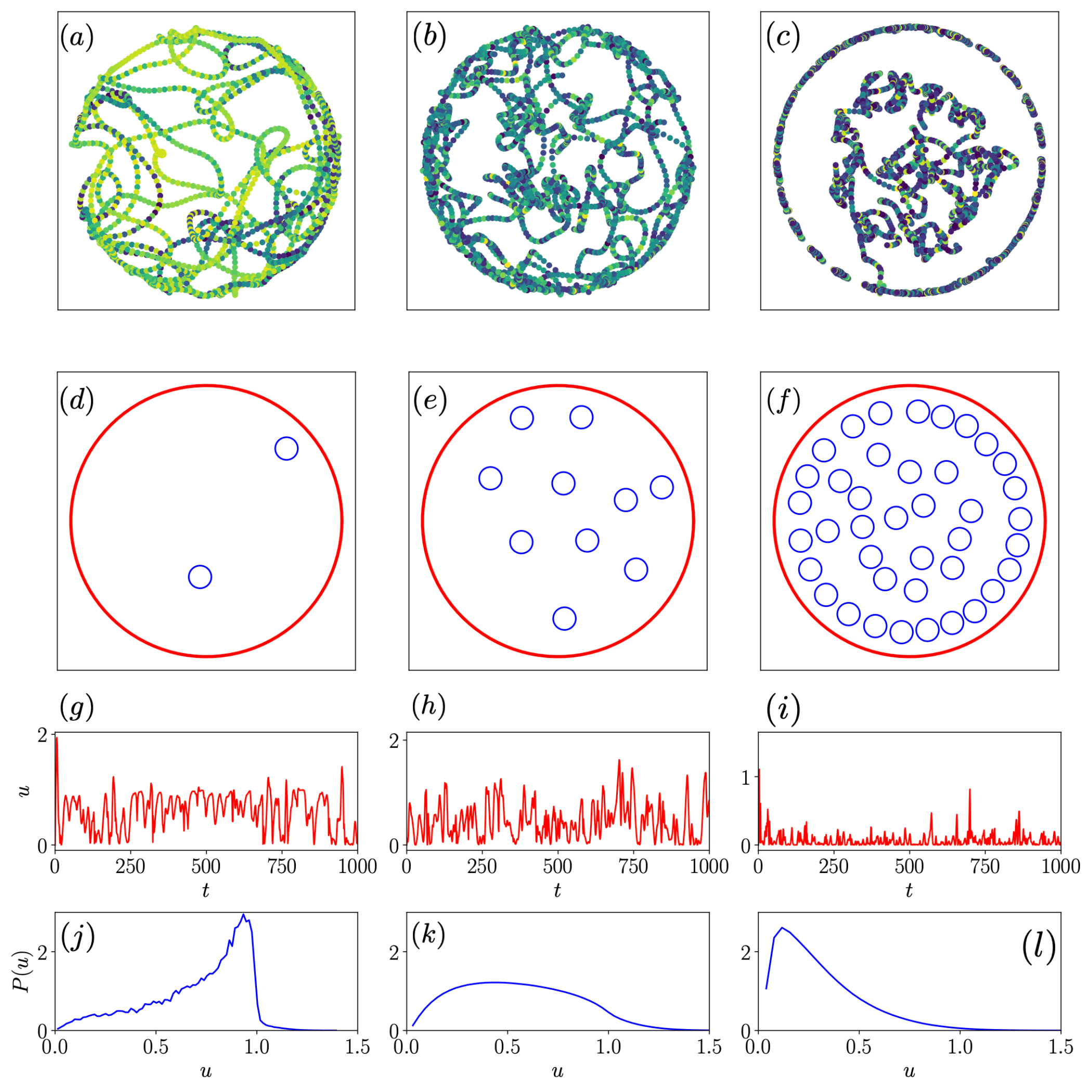} 
       \caption{ {\bf Representative trajectories.}
(a)-(c) Trajectories from numerical simulations ($\phi=0.01,0.07,0.28$, from left to right).  Color code for speed increases from dark to light green. (d)-(f) Corresponding snapshots taken at the end of numerical simulations. (g)-(i) Trajectory of the average velocity $u(t)$ ($\phi=0.01,0.07,0.28$, from left to right). (j)-(l) Probability distribution function of velocity ($\phi=0.01,0.07,0.28$, from left to right) .
}
\label{fig:sim0}
   \end{figure}

We fixed $\epsilon_s = 5$ and examined the stationary dynamics as the packing fraction $\phi$ increased. Fig.~\ref{fig:sim0} shows representative trajectories, color-coded for speed, for $\phi = 0.01, 0.07, 0.28$ in panels (a–c), along with snapshots of the corresponding steady-state configurations in panels (d–f). As in the experiments, trajectories become increasingly localized at higher densities. The velocity magnitude $u(t) \!\equiv\! |\mathbf{u}(t)|$ also follows the decreasing trend shown in panels (g–i), and the velocity distribution evolves with $\phi$, showing a change in skewness in panels (j–l), consistent with experimental observations.

\begin{figure*}[ht!]
       \centering
\includegraphics[width=\textwidth]{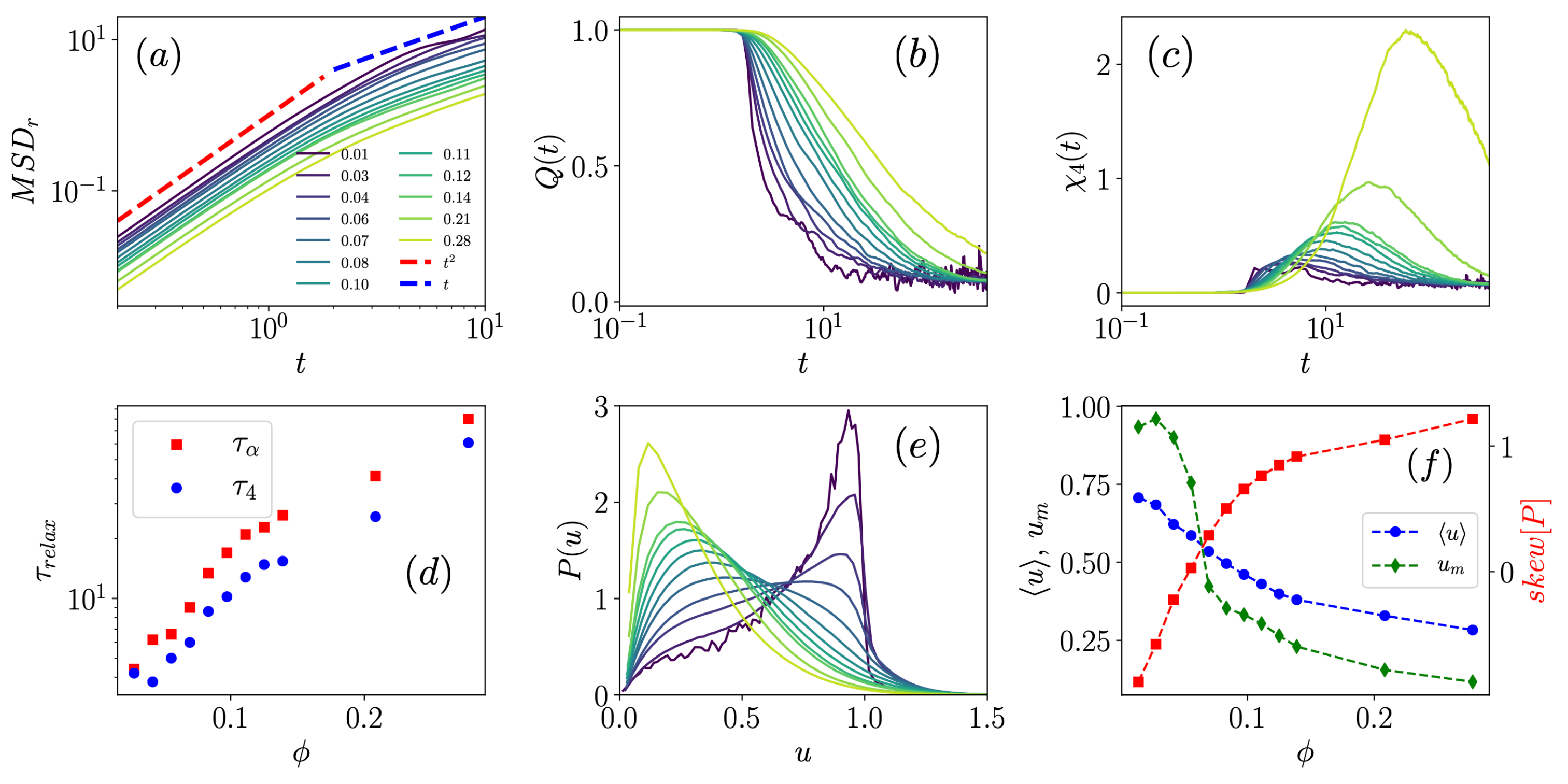}
       \caption{Results from numerical simulations. (a) $MSD_r$ as density increases. (b) Dynamical Overlap $Q(t)$. (c) Dynamical susceptibility $\chi_4(t)$. (d) Relaxation time as a function of density. (e) Probability distribution function of velocity. (f) Average velocity $\langle u \rangle$, $u_m$ (the value of $u$ that maximizes $P(u)$, i.e., the position of the peak of $P(u)$), and skewness of $P(u)$, as a function of $\phi$. }
       \label{fig:sim1}
\end{figure*} 

To quantify the dynamics, we measured key observables: the mean squared displacement $MSD_r$, the dynamical overlap $Q(t)$, and the velocity distribution $P(v)$, averaged over $N_s = 9750$ independent samples. From these, we also computed the dynamical susceptibility $\chi_4(t)$, which captures sample-to-sample fluctuations in $Q(t)$~\cite{lavcevic2003spatially}.
Fig.~\ref{fig:sim1}(a) shows that $MSD_r$ becomes increasingly subdiffusive at intermediate times as the packing fraction grows. This subdiffusive regime signals the onset of caging. The same effect is reflected in the dynamical overlap $Q(t)$ (Fig.~\ref{fig:sim1}(b)), which shows slower relaxation with increasing density.
The phase diagram already suggested the presence of dynamical heterogeneity, Fig.~\ref{fig:Numerics}(d). This is confirmed quantitatively by $\chi_4(t)$ (Fig.~\ref{fig:sim1}(c)), which shows the characteristic broad peak associated with heterogeneous relaxation in glassy systems. Notably, this behavior arises at unusually low densities, made possible by the second repulsive length scale in our model—unlike in standard active glass models. We quantified the relaxation time in two ways: (i) $\tau_\alpha$, defined by $Q(\tau_\alpha) = e^{-1}$, and (ii) $\tau_4$, the time at which $\chi_4(t)$ reaches its maximum. Both measures increase with $\phi$ (Fig.~\ref{fig:sim1}(d)), demonstrating consistent dynamical slowing down.

Consistent with the experiments, the velocity distribution $P(v)$ changes its skewness from negative to positive as density increases, Fig.~\ref{fig:sim1}(e). Finally, the slowing down of the dynamics is also evident in the average velocity, which decreases with increasing density (Fig.~\ref{fig:sim1}(f)). Examining different observables—(i) the mean velocity $\langle u \rangle$, (ii) the most probable velocity $u_m$, and (iii) the skewness—we identify a crossover around $\phi^* \approx 0.1$, where the skewness switches sign. This crossover coincides with the onset of strong localization and glassy dynamics.

\section{Conclusions and perspectives}

In this work, we studied the dynamics of a few-body system of confined camphor surfers—a minimal realization of macroscopic active matter. At intermediate particle densities, we observed intermittent bursting motion, characterized by abrupt collective rearrangements separated by quiescent periods. This regime coexists with glass-like features such as dynamical slowing and heterogeneity. Bursts become rarer with increasing density, a trend qualitatively captured by our analytic model that considers only minimal hydrodynamic coupling. Numerical simulations reproduced the glassy steady states—subdiffusion, caging, and heterogeneous dynamics—but not the bursting, which remains a target for future work. More broadly, our findings show that confinement and long-range interactions alone can generate complex temporal organization, even in the absence of alignment, motility-induced phase separation, or large system sizes. Few-body active systems thus provide a powerful platform for probing emergent dynamics at the boundary between single-particle and collective behavior.

\section*{Acknowledgements:} This work was partially supported by the National Science Foundation under grant no.~NSF DMS-2010018 and the Agence Nationale de Recherche under grant no.~ANR-23-CPJ1-0170-01 to WWA.  WWA also acknowledges funding and support from the Erskine Fellowship Program at the University of Canterbury. MP acknowledges funding from the Italian Ministero dell’Università e della Ricerca under the programme PRIN 2022 (``re-ranking of the final lists''), number 2022KWTEB7, cup B53C24006470006.
ML acknowledges the use of computational resources from the “Mésocentre” computing center of Université Paris-Saclay, CentraleSupélec and École Normale Supérieure Paris-Saclay supported by CNRS and Région Île-de-France (https://mesocentre.universite-paris-saclay.fr/). 

\bibliography{mpbib}
\bibliographystyle{rsc}

\end{document}